\documentclass{jps-cp}
\usepackage{txfonts} 

\title{Tracing Monopoles and Anti-monopoles in a Magnetic Hedgehog Lattice}

\author{Shun~\textsc{Okumura}$^1$, Satoru~\textsc{Hayami}$^2$, Yasuyuki~\textsc{Kato}$^1$, and Yukitoshi~\textsc{Motome}$^1$}

\inst{$^1$Department of Applied Physics, the University of Tokyo, Tokyo 113-8656, Japan\\
$^2$Department of Physics, Hokkaido University, Sapporo 060-0810, Japan}

\email{s.okumura@aion.t.u-tokyo.ac.jp}

\abst{
The magnetic hedgehog lattice (HL), which was recently discovered in the $B$20-type chiral magnet MnSi$_{1-x}$Ge$_x$, is a topological spin texture with a periodic array of magnetic monopoles and anti-monopoles.
Within the continuum approximation, the monopoles and anti-monopoles are predicted to move, collide, and pair annihilate in an applied magnetic field, but it remains unclear how the lattice discretization affects their motions. 
Here, we study the trajectories of monopoles and anti-monopoles in a lattice system by simulated annealing with field sweep.
We show that the monopoles and anti-monopoles move and repel before pair annihilations.
We also clarify that their motions are closely related with the field dependence of the scalar spin chirality.
}

\kword{topological spin textures, magnetic hedgehog lattice, topological Hall effect}

\begin{document}
\maketitle

\section{Introduction}
\begin{figure}[b]
\centering
\includegraphics[width=\columnwidth,clip]{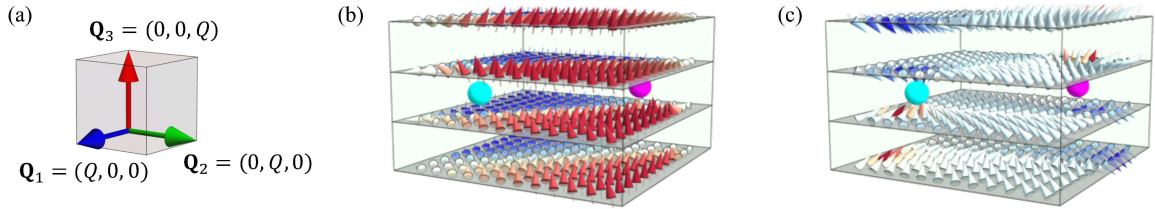}
\caption{
Schematic pictures of (a) the three ordering vectors, (b) the spin texture, and (c) the effective magnetic field of the 3$Q$ hedgehog lattice.
In (b) and (c), we plot four planes near a pair of hedgehog and anti-hedgehog.
The magenta (cyan) ball represents the position of the hedgehog (anti-hedgehog) core at an interstitial position.
The color of arrows in (b) and (c) represents the $z$ component of the magnetic moments and the effective magnetic field, respectively. 
(c) indicates that the hedgehog and anti-hedgehog are a monopole and anti-monopole of the effective field, respectively.
}
\label{f1}
\end{figure}

Recently, three-dimensional topological spin textures were discovered in the $B$20-type compounds MnSi$_{1-x}$Ge$_x$~\cite{Kanazawa2012,Tanigaki2015,Fujishiro2019}, which are called the magnetic hedgehog lattices (HLs). 
One example found in MnGe is the $3Q$-HL, which is characterized by cubic three ordering vectors~\cite{Kanazawa2012,Tanigaki2015} [Fig.~\ref{f1}(a)].
The $3Q$-HL has a periodic array of topological defects called hyperbolic hedgehogs and anti-hedgehogs [Fig.~\ref{f1}(b)], which generate effective magnetic fields regarded as monopoles and anti-monopoles, respectively~\cite{Kanazawa2016} [Fig.~\ref{f1}(c)].
The monopoles and anti-monopoles are connected with each other by fictitious fluxes, which are closely related to the topological Hall effect~\cite{Kanazawa2011,Kanazawa2012}. 

The monopoles and anti-monopoles are moved by an applied magnetic field, associated with the modulation of the spin texture.
Their motions were theoretically studied by the continuum approximation, and the shift, collision, and pair annihilation were predicted while increasing the field~\cite{Zhang2016}.
In MnGe, however, the $3Q$-HL has a very short period of $\sim 3$~nm, which is far from the continuum limit.
Hence, for understanding of the intriguing properties in real compounds, it is worth elucidating how the lattice discretization affects the positions of monopoles and anti-monopoles.

In this report, we study the trajectories of monopoles and anti-monopoles for the effective spin lattice model that reproduces short-period HLs~\cite{Okumura2019}.
By simulated annealing with field sweep, we clarify how the monopoles and anti-monopoles in a $3Q$-HL move while increasing the field.
We also discuss how their motions affect the uniform scalar spin chirality via the fictitious fluxes.

\section{Model and method}
\subsection{Model}
In the present study, we consider an effective spin model including contributions from both the spin-orbit and spin-charge couplings, following the previous study~\cite{Okumura2019}. 
The Hamiltonian is given by 
\begin{align}
\mathcal{H} = \sum_{\eta}\Big[-J\mathbf{S}_{\mathbf{Q}_\eta}\cdot\mathbf{S}_{-\mathbf{Q}_\eta}+\frac{K}{N}({\mathbf S}_{\mathbf{Q}_\eta}\cdot{\mathbf S}_{-\mathbf{Q}_\eta})^2-i{\mathbf D}_\eta\cdot{\mathbf S}_{\mathbf{Q}_\eta}\times{\mathbf S}_{-\mathbf{Q}_\eta}\Big]-h\sum_{l}S^z(\mathbf{r}_l).
\label{eq:Heff}
\end{align}
The first and second terms represent the Ruderman-Kittel-Kasuya-Yosida interaction~\cite{Ruderman1954,Kasuya1956,Yosida1957} and the biquadratic interaction with a positive coupling constant $K>0$~\cite{Akagi2012,Hayami2014,Hayami2017}, respectively, both of which are derived from the perturbation expansion with respect to the spin-charge coupling. 
Meanwhile, the third term describes the Dzyaloshinskii-Moriya type interaction~\cite{Dzyaloshinskii1958,Moriya1960,Hayami2018}, arising from the spin-orbit coupling. 
The sum of $\eta$ is taken for a set of the wave vectors $\mathbf{Q}_\eta$, and $\mathbf{S}_{\mathbf{q}} = \frac{1}{\sqrt{N}} \sum_l \mathbf{S}(\mathbf{r}_l)e^{i\mathbf{q}\cdot\mathbf{r}_l}$, where $\mathbf{S}(\mathbf{r}_l)=(S^x(\mathbf{r}_l),S^y(\mathbf{r}_l),S^z(\mathbf{r}_l))$ repersents the spin at site $l$, $\mathbf{r}_l$ is the position vector of the site $l$, and $N$ is the number of spins. 
The last term describes the Zeeman coupling to an external magnetic field $h$; here we consider the field along the [001] direction. 

In the following calculations, corresponding to the $3Q$-HL, we assume a set of the cubic wave vectors as $\mathbf{Q}_1=(Q,0,0)$, $\mathbf{Q}_2=(0,Q,0)$, and $\mathbf{Q}_3=(0,0,Q)$ with $Q=\pi/12$ [Fig.~\ref{f1}(a)], and $\mathbf{D}_\eta\parallel\mathbf{Q}_\eta$. 
We consider a simple cubic lattice with $N=24^3$, which matches the size of the magnetic unit cell set by $\mathbf{Q}_\eta$, under periodic boundary conditions and treat the spins as classical vectors with the length $|\mathbf{S}(\mathbf{r}_l)|=1$ for simplicity. 
We set the energy scale as $J=1$ and take $K=0.7$ and $D=0.3$ so as to stabilize the $3Q$-HL at zero field~\cite{Okumura2019}.

\subsection{Simulated annealing and field sweep}
\label{sec:annealing}
To obtain the spin texture realized for the model in Eq.~(\ref{eq:Heff}), we adopt simulated annealing with a magnetic field sweep.
We first perform the simulated annealing at zero field by gradually reducing the temperature from $T=1$ to $T\simeq10^{-5}$ with scheduling $T_n=10^{-0.1n}$, where $T_n$ is the temperature in the $n$th step.
During the annealing, we spend a total of $10^5$--$10^6$ Monte Carlo (MC) sweeps by using the standard Metropolis algorithm.
After obtaining the stable spin state at zero field, we sweep the field $h$ from $h=0$ to $1$ successively by $\Delta h=0.01$.
At every shift by $\Delta h$, we raise the temperature to $T\simeq10^{-3}$ for the spin configuration obtained in the previous annealing, and perform the simulated annealing down to $T\simeq10^{-5}$ with the same scheduling. 
Note that in this method we may follow a metastable state beyond the first-order phase transitions. 
Indeed, as shown in Sec.~\ref{sec:field}, the phase boundary is different from that in the stable ground state obtained in Ref.~\cite{Okumura2019}.
Nevertheless, we adopt the field sweep to trace the motions of monopoles and anti-monopoles beyond the phase transitions for capturing their pair annihilation process.

\subsection{Detection of monopoles and anti-monopoles}
\begin{figure}[t]
\centering
\includegraphics[width=\columnwidth,clip]{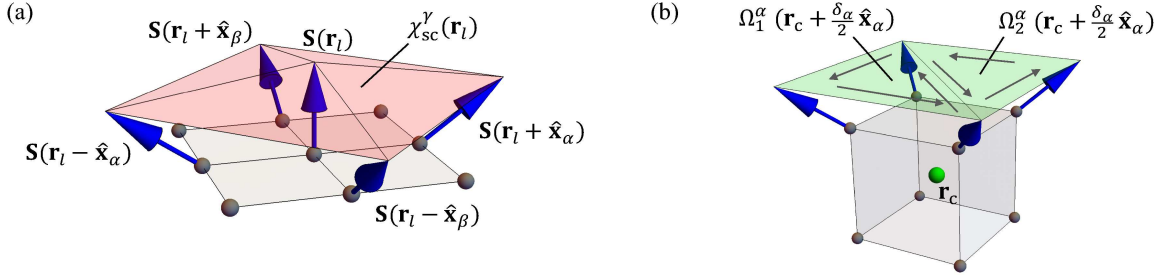}
\caption{
Schematic pictures of (a) the local scalar spin chirality $\chi^\gamma_\mathrm{sc}(\mathbf{r}_l)$ in Eq.~(\ref{eq:sc}) and (b) the solid angle $\Omega^\alpha(\mathbf{r}_\mathrm{c}+\frac{\delta_\alpha}{2}\hat{\mathbf{x}}_\alpha)$ in Eq.~(\ref{eq:Omega}).
The gray spheres represent the lattice sites on the cubic lattice, and the blue arrows are the spins at each site.
The red and green triangles represent the local scalar spin chirality and the solid angle, respectively.
In (b), the gray arrows denote the order of the outer products in Eq.~(\ref{eq:sa}) and the green ball represents the center of the unit cube. 
}
\label{f2}
\end{figure}

In order to detect the positions of monopoles and anti-monopoles for the spin configurations obtained by the simulated annealing, we calculate the local scalar spin chirality and the monopole charge. 
We define the local scalar spin chirality at a lattice site $\bold{r}_l$ in a vector form of $\boldsymbol{\chi}_{\rm sc}(\bold{r}_l) = (\chi^x_\mathrm{sc}(\mathbf{r}_l), \chi^y_\mathrm{sc}(\mathbf{r}_l), \chi^z_\mathrm{sc}(\mathbf{r}_l))$;
the $\gamma=x,y,z$ component is defined by using the spin at $\mathbf{r}_l$ and four neighboring spins in the plane perpendicular to the $\gamma$ direction, as shown in Fig.~\ref{f2}(a), whose explicit form is given by 
\begin{align}
\chi^\gamma_\mathrm{sc}(\mathbf{r}_l)=\frac{1}{2}\sum_{\alpha\beta\delta_\alpha\delta_\beta}\epsilon^{\alpha\beta\gamma}\delta_\alpha\delta_\beta\mathbf{S}(\mathbf{r}_l)\cdot\left[\mathbf{S}(\mathbf{r}_l+\delta_\alpha\hat{\mathbf{x}}_\alpha)\times\mathbf{S}(\mathbf{r}_l+\delta_\beta\hat{\mathbf{x}}_\beta)\right],
\label{eq:sc}
\end{align}
where $\delta_\alpha=\pm1$, $\hat{\mathbf{x}}_\alpha$ is the unit translation vector in the $\alpha=x,y,z$ direction, and $\epsilon^{\alpha\beta\gamma}$ is the Levi-Civita symbol.

We also define the monopole charge in each unit cube by using the solid angles of eight spins at the vertices. 
We compute the solid angle on a square surface of the unit cube perpendicular to the $\alpha$ direction by the sum of the solid angles of three spins on two triangles [Fig.~\ref{f2}(b)],
\begin{align}
\Omega^\alpha(\mathbf{r}_\mathrm{c}+\frac{\delta_\alpha}{2}\hat{\mathbf{x}}_\alpha)=\Omega^
\alpha_1(\mathbf{r}_\mathrm{c}+\frac{\delta_\alpha}{2}\hat{\mathbf{x}}_\alpha)+\Omega^\alpha_2(\mathbf{r}_\mathrm{c}+\frac{\delta_\alpha}{2}\hat{\mathbf{x}}_\alpha),
\label{eq:Omega}
\end{align}
where $\mathbf{r}_\mathrm{c}$ denotes the center of the unit cube; $\Omega^\alpha_i$ is defined by~\cite{Yang2016} 
\begin{align}
\Omega^\alpha_i(\mathbf{r}_\mathrm{c}+\frac{\delta_\alpha}{2}\hat{\mathbf{x}}_\alpha)= 2\tan^{-1}\left(\frac{\mathbf{S}_1\cdot\left[\mathbf{S}_2\times\mathbf{S}_3\right]}{|\mathbf{S}_1||\mathbf{S}_2||\mathbf{S}_3|+\mathbf{S}_1\cdot\mathbf{S}_2|\mathbf{S}_3|+\mathbf{S}_2\cdot\mathbf{S}_3|\mathbf{S}_1|+\mathbf{S}_3\cdot\mathbf{S}_1|\mathbf{S}_2|}\right).
\label{eq:sa}
\end{align}
where $\mathbf{S}_1$, $\mathbf{S}_2$, amd $\mathbf{S}_3$ are the three spins on the triangle. 
Note that the order of the outer product is taken in the order denoted by the gray arrows in Fig.~\ref{f2}(b), and the sign of $\Omega^\alpha_i$ is taken to be the same as that of $\mathbf{S}_1\cdot\left[\mathbf{S}_2\times\mathbf{S}_3\right]$: $\Omega^\alpha_i\in[-2\pi,2\pi]$.
By using the solid angles, we calculate the monopole charge $Q_\mathrm{m}$ as 
\begin{align}
Q_\mathrm{m}(\mathbf{r}_\mathrm{c}) = \frac{1}{4\pi}\sum_{\alpha\delta_\alpha}\delta_\alpha\Omega^\alpha(\mathbf{r}_\mathrm{c}+\frac{\delta_\alpha}{2}\hat{\mathbf{x}}_\alpha),
\label{eq:monopole_number}
\end{align}
which takes the value of +1 (-1) when a monopole (anti-monopole) exists in the unit cube.

\subsection{Other physical quantities}
In addition, we calculate the magnetization per site along the [001] direction, $m=\frac{1}{N}\sum_{l}S^z(\mathbf{r}_{l})$, and the uniform scalar spin chirality per site, 
$\chi_\mathrm{sc}=\frac{1}{N}\sum_l\chi^z_\mathrm{sc}(\mathbf{r}_l)$.
We also compute the total number of monopoles and anti-monopoles in the magnetic unit cell, $N_\mathrm{m}=\sum_{\mathbf{r}_\mathrm{c}\in\mathrm{unit \, cell}}|Q_\mathrm{m}(\mathbf{r}_\mathrm{c})|$.
In addition, we calculate the magnetic moment with wave vector $\mathbf{q}$, $m_\mathbf{q}=\sqrt{S(\mathbf{q})/N}$, where $S(\mathbf{q})$ is the spin structure factor defined by $S(\mathbf{q})=\frac{1}{N}\sum_{ll'}\mathbf{S}(\mathbf{r}_l)\cdot\mathbf{S}(\mathbf{r}_{l'})e^{i\mathbf{q}\cdot(\mathbf{r}_l-\mathbf{r}_{l'})}$.

\section{Results}
\subsection{Zero magnetic field}
\begin{figure}[h]
\centering
\includegraphics[width=\columnwidth,clip]{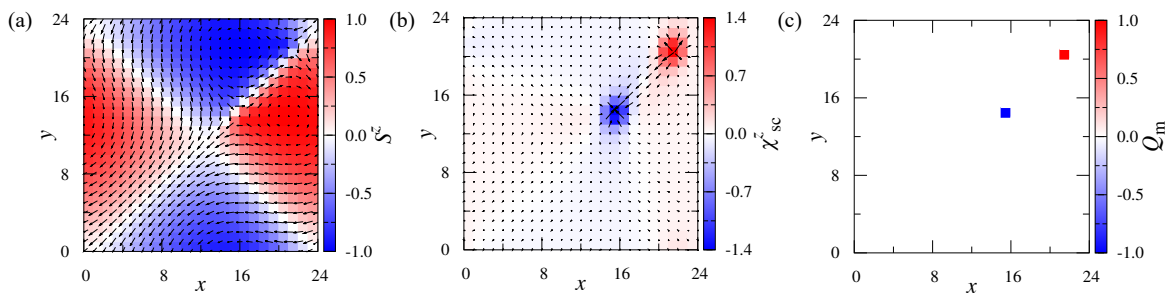}
\caption{
MC snapshot on an $xy$ plane after the simulated annealing for the model in Eq.~(\ref{eq:Heff}) at zero field: (a) the spin texture $\mathbf{S}$, (b) the scalar spin chirality $\boldsymbol{\chi}_\mathrm{sc}=(\chi^x_\mathrm{sc},\chi^y_\mathrm{sc},\chi^z_\mathrm{sc})$ [Eq.~(\ref{eq:sc})], and (c) the monopole charge $Q_\mathrm{m}$ [Eq.~(\ref{eq:monopole_number})].
(c) is the plot on the plane lower in the $z$ direction by half of the lattice constant to that in (a) and (b). 
In (a) and (b), the black arrows denote the $x$ and $y$ components and the colors represent the $z$ components.
}
\label{f3}
\end{figure}

First, we show the results at zero field, which is essentially the same as those obtained in Ref.~\cite{Okumura2019}.
Figure~\ref{f3}(a) displays a MC snapshot on an $xy$ plane after the simulated annealing.
The spin configuration is a noncoplanar one similar to Fig.~\ref{f1}(b), suggesting the existence of a pair of magnetic hedgehog and anti-hedgehog near the plane.
Figure~\ref{f3}(b) shows the local scalar spin chirality $\boldsymbol{\chi}_{\rm sc}(\bold{r}_l)$ computed by Eq.~(\ref{eq:sc}) using the spin texture in Fig.~\ref{f3}(a). 
As shown in the figure, $|\chi^z_\mathrm{sc}|$ takes large absolute values around two points, which suggest the hedgehog and anti-hedgehog. 
Figure~\ref{f3}(c) shows the monopole charge $Q_\mathrm{m}(\mathbf{r}_\mathrm{c})$ calculated by Eq.~(\ref{eq:monopole_number}). 
$Q_\mathrm{m}$ takes $+1$ and $-1$ at the locations with large $|\chi_{\rm sc}^z|$, indicating that the monopole and anti-monopole are present at the cores of the magnetic hedgehog and anti-hedgehog at the interstitial positions adjacent to the $xy$ plane.
Note that $Q_\mathrm{m}$ is useful to identify the positions of monopoles and anti-monopoles more accurately than $\boldsymbol{\chi}_{\rm sc}$.
By repeating the analysis to all the $xy$ planes, we find similar four pairs of monopoles and anti-monopoles in the magnetic unit cell, which form a spiral in the [100], [010], and [001] directions (see Fig.~\ref{f5}).
As shown in Fig.~\ref{f3}(b), the monopole and anti-monopole are the source and sink of $\boldsymbol{\chi}_{\rm sc}(\bold{r}_l)$, respectively; the flow of $\boldsymbol{\chi}_{\rm sc}(\bold{r}_l)$ between the monopole and anti-monopole is called the fictitious flux.

\subsection{In a magnetic field}
\label{sec:field}
\begin{figure}[t]
\centering
\includegraphics[width=\columnwidth,clip]{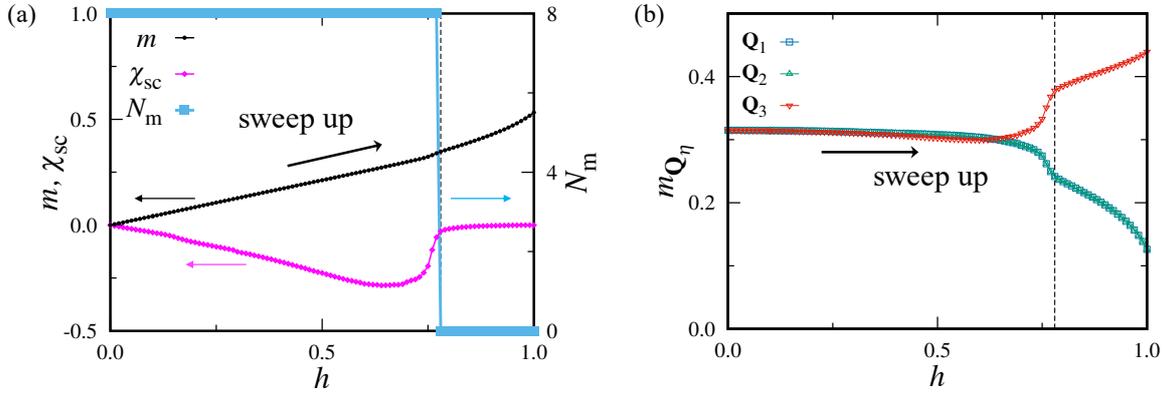}
\caption{
Magnetic field sweep along the [001] direction, starting from the $3Q$-HL at zero field: (a) the magnetization $m$, the uniform scalar spin chirality $\chi_{\rm sc}$, the number of monopoles and anti-monopoles $N_{\rm m}$, and (b) the magnetic moments with wave vector $\mathbf{Q}_\eta$, $m_{\mathbf{Q}_\eta}$. 
The results are calculated for $D=0.3$ and $K=0.7$. 
The dashed line shows the critical magnetic field where $N_\mathrm{m}$ vanishes.
}
\label{f4}
\end{figure}

Next, we show the results obtained by sweeping up the magnetic field along the [001] direction.
As shown in Fig.~\ref{f4}(a), the magnetization $m$ smoothly increases as increasing $h$, and shows a kink at $h\simeq0.78$, where the number of the monopoles and anti-monopoles $N_\mathrm{m}$ vanishes.
The uniform scalar spin chirality in Fig.~\ref{f4}(a) is also reduced rapidly around $h\simeq0.78$.
For both below and above $h\simeq0.78$, $m_{\mathbf{Q}_\eta}$ is nonzero for three $\mathbf{Q}_\eta$, as shown in Fig.~\ref{f4}(b).
Therefore, the results indicate that the $3Q$-HL turns into a topologically-trivial $3Q$ state at $h\simeq0.78$.
Note that we follow a metastable state by the field sweep, as mentioned in Sec.~\ref{sec:annealing}; the true phase diagram is more complicated with several first-order transitions between different $3Q$ states~\cite{Okumura2019}. 

\begin{figure}[t]
\centering
\includegraphics[width=\columnwidth,clip]{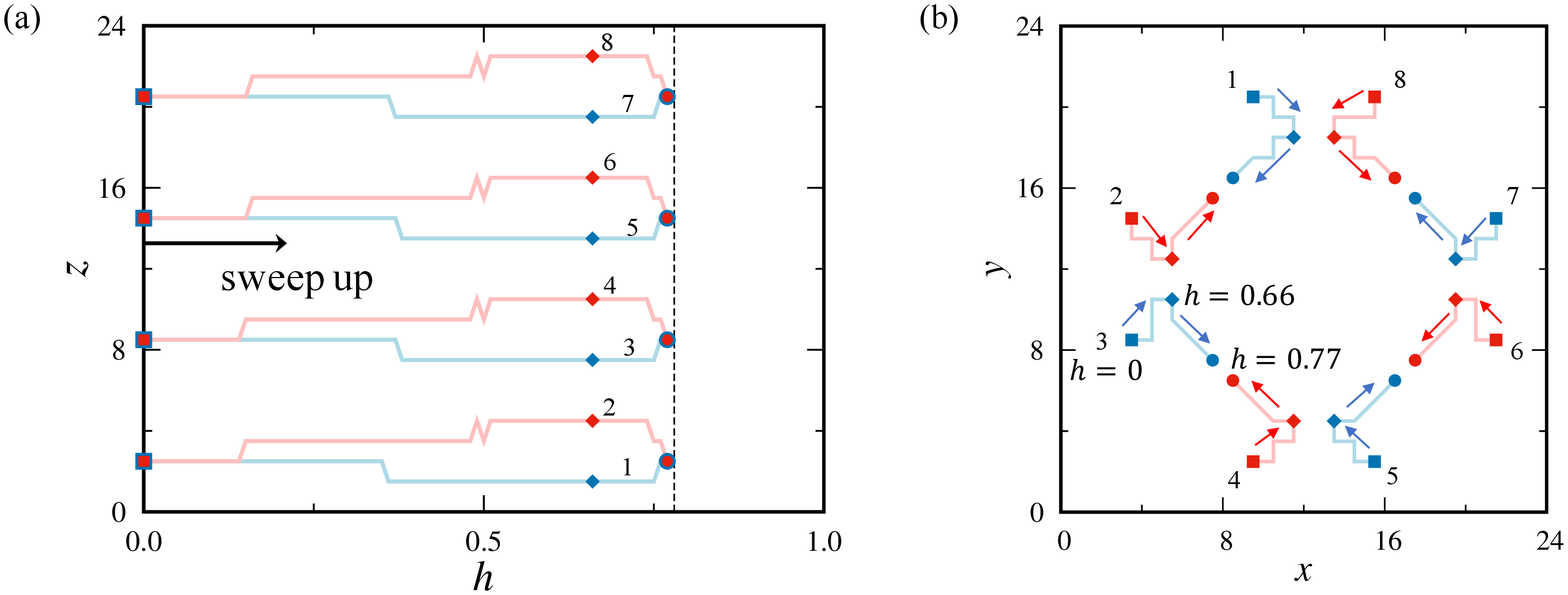}
\caption{
Trajectories of monopoles and anti-monopoles in the [001] field: (a) the $z$ positions as functions of the magnetic field $h$ and (b) the projection onto the $xy$ plane.
The red (blue) lines represent the trajectories of monopoles (anti-monopoles).
The solid squares, diamonds, and circles denote the positions at $h=0$, 0.66, and 0.77, respectively.
The dashed line in (a) shows the critical magnetic field by the pair annihilation of the monopoles and anti-monopoles.
}
\label{f5}
\end{figure}

Figure~\ref{f5} shows the trajectories of the monopoles and anti-monopoles in this field sweep, drawn by tracing the positions where $Q_{\rm m} = \pm1$.
At $h=0$, there are four pairs of monopoles and anti-monopoles, (1,2), (3,4), (5,6), and (7,8), on the $xy$ planes with $z=2.5$, $8.5$, $14.5$, and $20.5$, respectively, as shown in Fig.~\ref{f5}(a). 
Their positions projected onto the $xy$ plane are shown by the squares in Fig.~\ref{f5}(b). 
Note that the pair shown in the Fig.~\ref{f3}(c) corresponds to (5,6).
When increasing the magnetic field, the (anti-)monopoles move to the upper (lower) layers, while they move in the $xy$ directions as well, as shown in Fig.~\ref{f5}. 
These shifts indicate that the fictitious fluxes between the pairs incline to the $z$ direction, leading to the increase of the absolute value of the uniform scalar spin chirality shown in Fig.~\ref{f4}(a).
At $h\simeq0.66$, the monopoles are repelled by anti-monopoles coming from different layers, as shown in Fig.~\ref{f5}(a); for instance, the monopole 4 is repelled by the anti-monopole 5.
Thereafter, the monopoles and anti-monopoles return to the initial $xy$ planes and annihilate with each other at $h\simeq0.78$.
This process rapidly reduces the uniform scalar spin chirality, as shown in Fig.~\ref{f4}(a).

\section{Concluding remarks}
Our analyses on the metastable $3Q$-HL in the field sweep clarified the motions of the topological defects, monopoles and anti-monopoles, on the discrete lattice.
Our results also explicitly show that how their trajectories are related with the uniform scalar spin chirality and the fictitious fluxes. 
One of the differences from the predictions by the continuum approximation~\cite{Zhang2016} is that the monopoles and anti-monopoles do not collide with each other but are repelled by other monopoles and anti-monopoles before pair annihilations. 
This suggests the existence of repulsive interactions between the monopoles and anti-monopoles coming from different layers, when the spin textures are optimized on the discrete lattice during the field sweep. 

Our findings provide a first step to understand the intriguing phenomena discovered in the short-period HLs in MnSi$_{1-x}$Ge$_x$, such as the topological Hall effect~\cite{Kanazawa2011} and the topological thermoelectric transport~\cite{Shiomi2013, Fujishiro2018}. 
Our calculations can be straightforwardly extended to other field directions and for other types of HLs, such as the $4Q$-HL~\cite{Fujishiro2019,Okumura2019}. 
For detailed comparisons with the experiments, however, it is desired to sophisticate the theory by including the realistic lattice structure and model parameters.
\\

This research was supported by JST CREST (No.~JPMJCR18T2) and the JSPS KAKENHI (No.~JP19H05825 and No.~JP18K13488).
This research was also supported by the Chirality Research Center in Hiroshima University and JSPS Core-to-Core Program, Advanced Research Networks.
S. O. was supported by JSPS through the research fellowship for young scientists.

\end{document}